# High-speed InAs/GaSb Mid-Wave Infrared Interband Cascade Photodetector at Room Temperature

Zhiyang Xie,[1,4,5,†] Jian Huang,[1,4,5,†] Xuliang Chai,[2,5,†] Zhuo Deng,[1] Yaojiang Chen,[1] Qi Lu,[6] Zhicheng Xu,[2,3] Jianxin Chen,[2,3] Yi Zhou,[2,3,*] and Baile Chen[1,*]

[1]School of Information Science and Technology, ShanghaiTech University, Shanghai 201210, China

[2]Key Laboratory of Infrared Imaging Materials and Detector, Shanghai Institute of Technical Physics, Chinese Academy of Sciences, Shanghai 200083, China

[3]Hangzhou Institute for Advanced Study, University of Chinese Academy of Sciences, No.1, Sub-Lane Xiangshan, Xihu District, 310024, Hangzhou, China

[4]Shanghai Institute of Microsystem and Information Technology, Chinese Academy of Sciences, Shanghai 200050, China

[5]University of Chinese Academy of Sciences, Beijing 100049, China

[6]Physics Department, Lancaster University, Lancaster, LA1 4YB, United Kingdom.

ABSTRACT: High speed mid-wave infrared (MWIR) photodetectors have important applications in the emerging areas such high-precision frequency comb spectroscopy and light detection and ranging (LIDAR). In this work, we report a high-speed room-temperature mid-wave infrared interband cascade photodetector (ICIP) based on a type-II InAs/GaSb superlattice. The devices show an optical cut-off wavelength around 5μm and a 3-dB bandwidth up to 7.04 GHz. The relatively low dark current density around $9.39 \times 10^{-2}$ A/cm$^2$ under −0.1 V is also demonstrated at 300K. These results validate the advantages of ICIPs to achieve both high-frequency operation and low noise at room temperature. Limitations on the high-speed performance of the detector are also discussed based on the S-parameter analysis and other RF performance measurement.

Semiconductor photodetectors sensitive to mid-wave infrared (MWIR) have attracted extensive attentions in many applications such as chemical sensing, gas monitoring and high-performance infrared imaging. In some special applications, such as free-space optical communication and frequency comb spectroscopy, MWIR photodetectors capable of high-frequency operation are required as an essential component[1-6]. Currently, quantum well infrared photodetectors (QWIPs) and quantum cascade photodetectors (QCDs) for high speed MWIR applications have been demonstrated[7-9]. Nevertheless performance of QCDs is mainly limited by absorption efficiency and noise due to a short carrier lifetime[10-15], and QWIPs have shortcomings in large dark currents and no response to normal incident light[7, 16].

Compared with conventional photodetectors, interband cascade infrared photodetectors (ICIPs) have more flexibility to alleviate the limitation on absorber thickness due to a finite diffusion length so that all photo-generated carriers can be efficiently collected, while absorption of incident light can be ensured with multiple absorbers located in each stage connected in series[16]. On the other hand, the individual absorber thickness in every stage can be shortened to reduce the carrier transit time across a single stage. Another advantage of ICIPs is that noise is suppressed by the multiple discrete short absorbers, instead of a single long absorber[17, 18]. As one of the most widely studied material system for MWIR, ICIPs based on InAs/GaSb type-II superlattice (T2SL) could offer several advantages such as wavelength tunability between 1 to 25 μm, excellent carrier transport properties and signal to noise ratio (SNR)[16]. Lotfi et al. reported a three-stage ICIP based on InAs/GaSb/AlSb/InSb T2SLs with a cutoff wavelength around 4.2μm at 300K, and the corresponding 3-dB bandwidth was ~1.3 GHz under zero bias[19]. Recently, a two-stage ICIP employing the InAs/Ga(As)Sb T2SLs as the absorption layer with cutoff wavelength of 5.6 μm and the 3-dB bandwidth of 2.4 GHz under −5 V bias at 300 K was demonstrated by our group[20].

In this paper, we report a five-stage ICIP based on InAs/GaSb type-II superlattice with a thin absorber of 240 nm in each stage. The dark current density is 1.04 A/cm$^2$ under −5 V bias at 300 K with cutoff wavelength of ~5 μm at 300 K. The 3-dB bandwidth of a 20 μm circular diameter detector achieves 7.04 GHz under −5 V at room temperature. Other RF performances of the detector such as saturation power are also characterized. The limitations on the high frequency performance characteristics are extensively studied based on the scattering parameter.

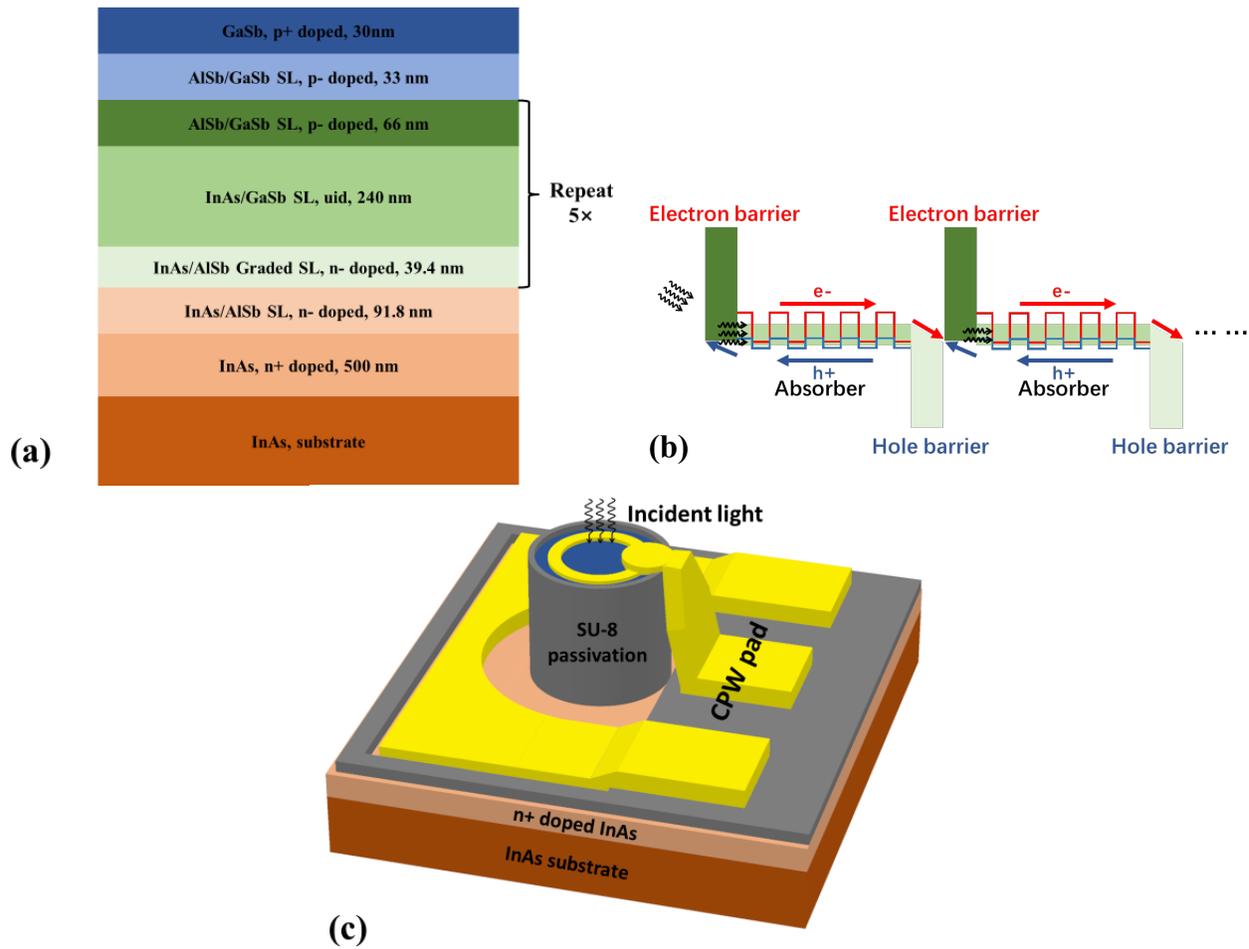

Figure 1. (a) Epitaxial structure of the designed five-stage ICIP. (b) Schematic diagram of the multiple-stage ICIP. The solid arrows show the movement of electrons/holes, and the dashed arrows represent the incident light. (c) Schematic diagram of the fabricated device.

**Device structure**

The epitaxial structure of the designed five-identical-stage ICIP is shown in Figure 1a. The sample was grown on InAs substrate by using molecular beam epitaxy system (MBE). The epitaxial growth began with a 500 nm thick n-type InAs bottom contact layer. Then, a 91.8 nm thick n-type InAs/AlSb superlattice (SL) was grown followed by a five-indentical-stage interband cascade structure using InAs/GaSb type-II SL as absorption layer. After that, a 33 nm thick p-type GaSb/AlSb SL was grown, and finally the structure was capped by a 30 nm p-type GaSb top contact layer. Each cascade stage consists of a 50 periods un-intentionally doped InAs/GaSb (2.4 nm/2.4 nm) type-II SL absorption layer and sandwiched a relaxation and a tunneling region, which help the photogenerated carriers transport to adjacent stages. The relaxation and tunneling region also act as hole and electron barriers, respectively, which can reduce dark current associated with the generation-recombination process. The electron barrier was designed with 5 periods AlSb/GaSb (1.8 nm/4.8 nm) SL and the hole barrier consists of graded InAs/AlSb SL. The whole structure was properly designed so that the photo-generated eletrons in absorption layer can relax through the graded InAs/AlSb SL transport region and recombine with the holes from the adjacent GaSb/AlSb electron barrier region, as shown in Figure 1b.

In order to investigate the electrical and optical performance of the designed ICIP device, the sample was fabricated into mesa-isolated devices using standard UV lithography and citric acid based wet etch ($C_6H_8O_7$:$H_3PO_4$:$H_2O_2$:$H_2O$=1:1:4:16). The etched surface was passivated by SU-8 to help suppressing the surface leakage current[21]. Ti/Pt/Au metal were deposited on top and bottom n+ doped InAs contact layers to form good ohmic contact by using electron beam evaporation.

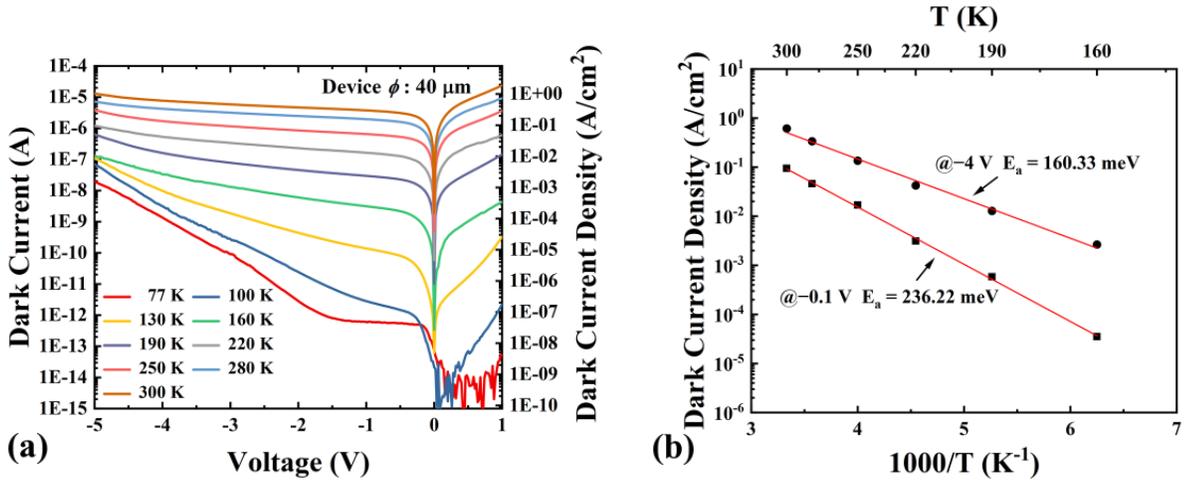

Figure 2. (a) Dark current voltage characteristics measured from the 40 μm diameter device at different temperatures from 77 to 300 K. (b) Arrhenius plot of the dark current density under −0.1 V and −4 V.

**Optical response and DC performance**

The temperature-dependent dark current-voltage characteristics of the ICIP device with 40 μm diameter at different temperatures from 77 to 300 K are shown in Figure 2a. The dark current was measured in a variable temperature probe station and recorded by a semiconductor device analyzer. The fluctuation and small photovaltaic shift of dark current around 0 V at low temperature are due to the imperfection of the cold shield and noise floor of the semiconductor device analyzer. The dark current density of the device varies from $2.35 \times 10^{-8}$ A/cm$^2$ to $9.39 \times 10^{-2}$ A/cm$^2$ under −0.1 V from 77 to 300 K.

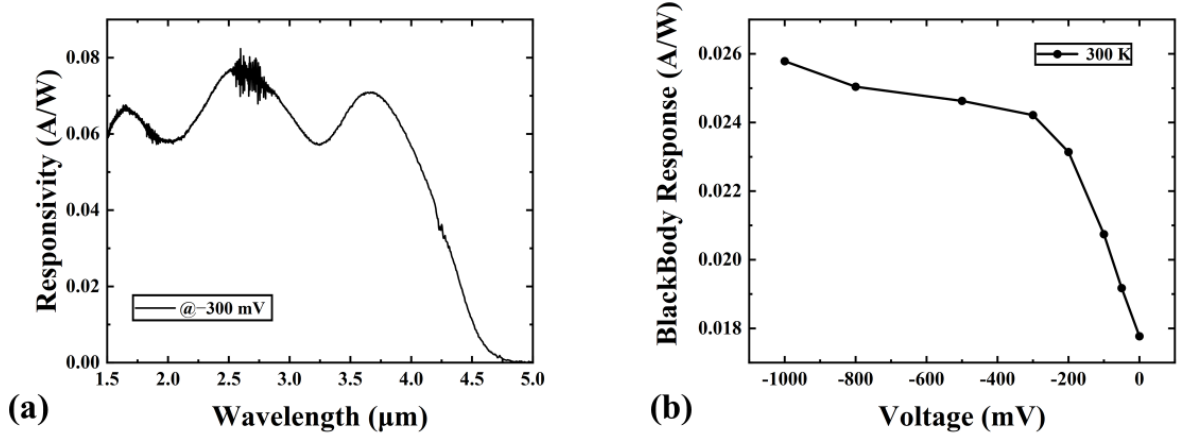

Figure 3. (a) Responsivity of the five-stage ICIP sample measured at room temperature under −0.3V bias. (b) Blackbody response of the device under different biases at 300K.

Compared with our previous work[22], this device shows a much lower dark current density. We believe this is due to more stages and thinner absorber used in this device. Figure 2b shows the Arrhenius plots of the device's dark current density as a function of temperature under −0.1 V and −4 V bias. The linear fits yield an activation energy of 236.2 meV under −0.1 V, which is very close to the effective bandgap (248.3 meV) of the five-stage ICIP, indicating that dark current is mainly dominated by diffusion component. Under −4 V bias, the activation energy decreases to 160.3 meV which implies the tunneling related processes begin to dominate at higher reverse bias. At 300 K, the dark current density increases slowly with reverse bias, and the device shows a dark current of 13.0 μA, corresponding to a dark current density of 1.04 A/cm$^2$ under −5 V bias.

After the electrical charaterization, the optical performance of the device was investigated. Figure 3a presents the responsivity of the five-stage ICIP device without anti-reflection coating at 300 K under −0.3 V bias voltage. The spectrum was measured by a Fourier transform infrared spectrometer (FTIR) and carlibrated by a blackbody source. As can be seen, the cutoff wavelength of the device is about 5 μm and the responsivty is about 0.067 A/W at 3.5 μm. The overall response of this device is smaller than that in previouly reported MWIR ICPDs[23], which mainly results from the much thinner absorption thickness. Figure 3b shows the blackbody responsivity as a function of reverse bias at 300 K of the device. Here, the blackbody responsivity (BBR) is defined as the ratio of the output photocurrent and the input radiation power of the blackbody source[24]. The BBR rapidly increases from 0.018 A/W to 0.024 A/W as the reverse bias varies from 0 V to −0.3 V. The reverse bias can help photogenic carriers overcome the barrier formed in the structure and also solve the problem of insufficient diffusion length of carriers at room temperature. With further increase of reverse bias, the BBR increases only slightly, suggesting the saturation of responsivity with reverse bias.

**Radio Frequency Characterization**

For RF measurement, normal-incident devices with 20 to 60 μm diameter were fabricated by UV photolithography and wet etching. Ti/Pt/Au were used for n-metal and p-metal contacts. The Ti layers provide good adhesion to GaSb and InAs, while the Pt layers prevent Au penetration into the semiconductor[25]. The devices were finally connected to a coplanar waveguide (CPW) pad of 50 Ω characteristic impedance through an air-bridge, which was electroplated on 2 μm thick SU-8 to achieve insulation of each device[20, 26], as shown in Figure 1c.

The frequency response of device at room temperature was investigated by a lightwave component analyzer (LCA) system. A lensed fiber was used to couple the modulated light of 1550 nm wavelength to the device. In order to achieve uniform illumination, the lensed fiber was lifted to the position where the responsivity dropped by about 50% during the measurement. DC bias was supplied by a source meter through one port of the bias tee, while the RF signal was collected by the LCA system through another port. The values of the frequency-dependent loss of cables and probe were carefully calibrated by Vector Network Analyzer.

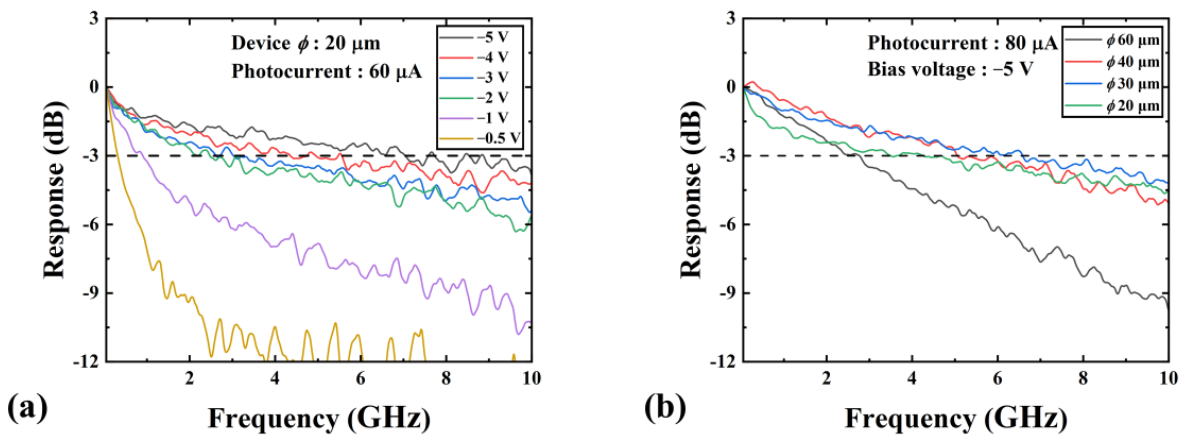

Figure 4. Room temperature frequency response of the five-stage ICIP (a) versus bias voltages with 20 μm device diameter under 60 μA photocurrent; (b) versus device diameter under 80 μA photocurrent and −5 V bias.

Figure 4a shows the result of the measured ICPD 3-dB bandwidth with 20 μm diameter as a function of bias voltages under 60 μA average photocurrent. The 20 μm diameter device achieves 0.91 GHz under −1 V bias and increases to 7.04 GHz when bias rises to −5 V. The 3-dB bandwidth of devices with different diameter (20 to 60 μm) under −5 V bias was also measured at 80 μA average photocurrent, as shown in Figure 4b. The 40 μm diameter device exhibits a bandwidth of 5.06 GHz under −5 V, which is higher than that InAs/Ga(As)Sb two-stage ICIP with the same diameter at 300 K (2.4 GHz) in our previous work[20]. This is due to the thinner absorber region used in this work, which could help reduce the carrier transit time and hence increase the 3-dB bandwidth. It is noted that the 20 μm diameter device has a 3-dB bandwidth of only 3.50 GHz with photocurrent of 80 μA, which is much lower than that with photocurrent of 60 μA due to the saturation of the device. As the optical power increases, the electric field in the depleted region is screened by the space charges and eventually collapses within the absorber region and thus the 3-dB bandwidth reduces. To further investigate the saturation characteristic of the five-stage ICIP, the RF output as a function of average photocurrent under various bias voltages was measured as shown in Figure 5. The saturation current is defined as the average photocurrent where the RF power compression curve drops by 1 dB from its peak value[5, 27]. Device with 40 μm diameter saturates at photocurrent of 290 μA at 5 GHz, while the 30 μm diameter device exhibits lower saturation photocurrent of 164 μA. The small saturation power could be limited by the multiple-stage architecture structure of the ICIP, which has significant current mismatch issue as more photo-generated carriers being created in the first stage.

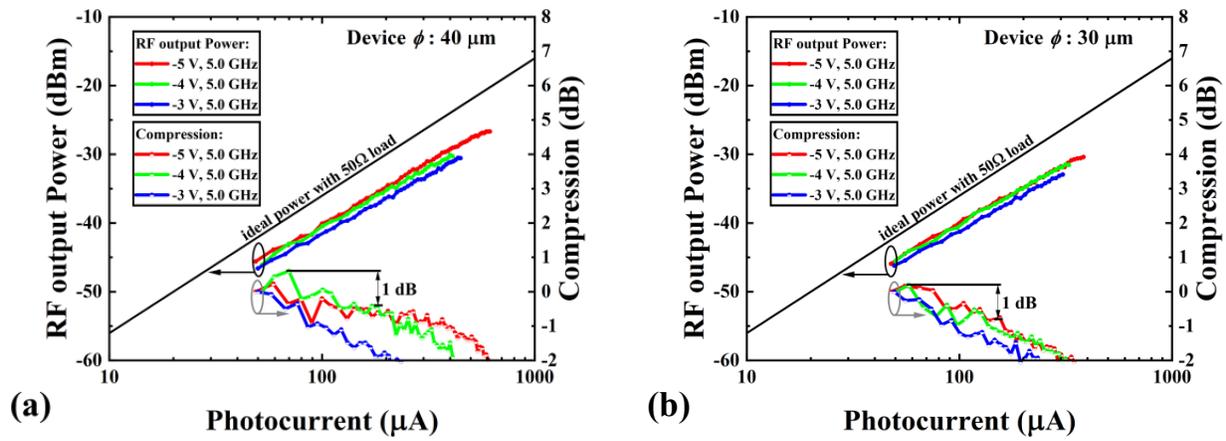

Figure 5. RF output power and RF power compression versus photocurrent under different bias at room temperature for device with (a) 40 μm diameter; (b) 30 μm diameter.

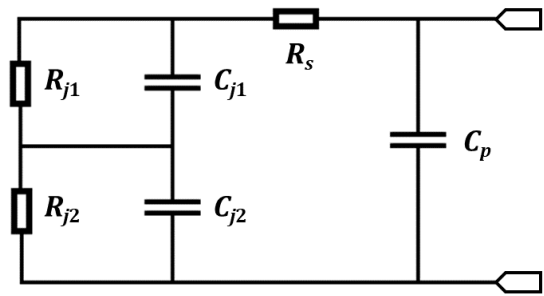

Figure 6. Equivalent circuit model of the ICIP for S11 fitting.

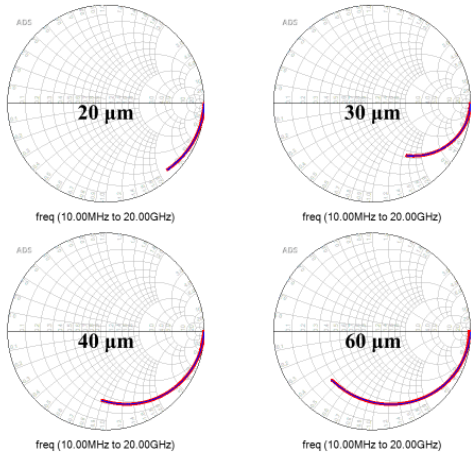

Figure 7. Measured and fitted (smooth line) S11 data with 10 MHz-20 GHz frequency range under −5 V bias voltage for ICIPs with various diameter at room temperature.

In order to study the bandwidth limiting factors of the device, the scattering parameter S11 of device was measured by using a Vector Network Analyzer. The parameter fitting was conducted in Advanced Design System (ADS) software with equivalent circuit model as depicted in Figure 6. The measured and fitted (smooth line) S11 data with 10 MHz-20 GHz frequency range of device under −5 V bias are shown in Figure 7. In the equivalent circuit model, $R_s$ represents the series resistance, and $C_p$ is the parasitic capacitance of the air-bridge and CPW pad. Similar to our previous work[20], the multiple-stage ICIP can also be modeled as two p-i-n junctions connecting in series in this circuit model, which includes junction resistance ($R_{j1}$,$R_{j2}$) paralleled with junction capacitance ($C_{j1}$,$C_{j2}$) in each series junction. The multiple number of stages in ICIP can be divided into two groups: depleted and un-depleted absorber under reverse bias. Apart from the depleted junctions which are represented as the first junction ($R_{j1}$ and $C_{j1}$), the other un-depleted stages are unaffected by external applied bias ($R_{j2}$ and $C_{j2}$). This model can avoid arbitrariness and complexity in the fitting process when using five series junctions to represent five stages in ICIP which will lead to ten variable parameters.

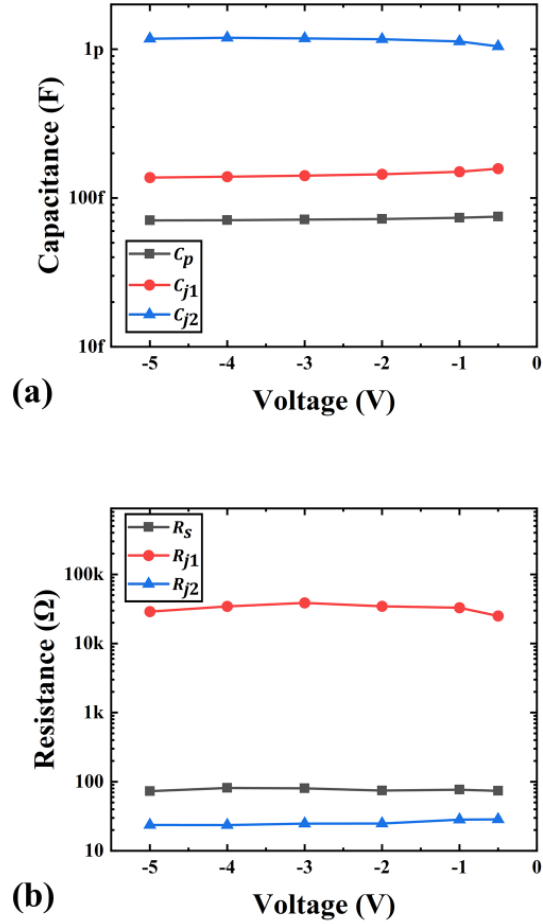

Figure 8. Dependence of extracted circuit parameters on bias voltage of the 40 μm five-stage ICIP at room temperature.

The extracted fitting results of the circuit model under various bias from −0.5 V to −5 V at room temperature are shown in Figure 8. By applying the simplified model of multiple-stage ICIP, the fitting results have similar trends as the two-stage ICIP in our previous work[20]. As expected, the values of parasitic capacitance ($C_p$) and series resistance ($R_s$) are independent of the bias voltage. In the first junction, the value of $R_{j1}$ increases from −0.5 V to −3 V due to the reduction of carrier density in depletion region, and then decreases from −3 V to −5 V, which can be attributed to the emergence of tunneling related current component as analyzed above. The decrease of $C_{j1}$ indicates the first junction is gradually depleted with the raise of reverse bias voltage, corresponding to the fully-depleted and partially-depleted junctions. The values of $C_{j2}$ and $R_{j2}$ show weak dependences on reverse bias, which also indicates these un-depleted stages are not significantly affected by external applied bias.

The thickness of depleted region can be approximately calculated by Eq. (1)

$$d = \frac{\varepsilon_0 \varepsilon_r A}{C} \quad (1)$$

where $\varepsilon_0$, $\varepsilon_r$, A, d, C is the permittivity of free space, the dielectric constant of absorber, area of device and width of depletion region, capacitance of photodiode respectively. Here the value of equivalent thickness ($d_{tot}$) of the first depleted junction is about 1.24 μm based on the capacitance value $C_{j1}$ at −5 V. Then, we estimate that only around three to four stages of ICIP are fully-depleted in series under −5 V bias. Thus, overall bandwidth of device could be limited by the slow carrier diffusion process in the un-depleted stages.

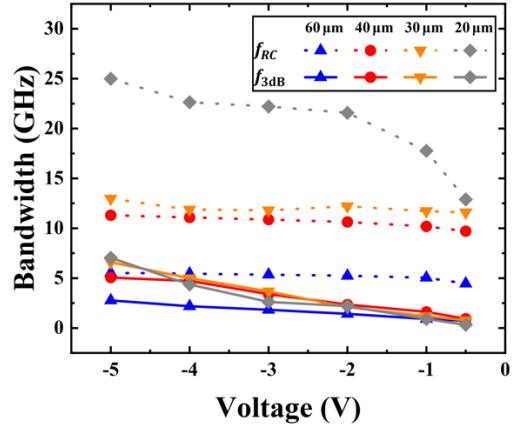

Figure 9. Bias voltage dependences of RC-limit bandwidth (dash lines) and measured 3-dB bandwidth (solid lines) of the five-stage ICIP under various bias voltage at room temperature.

From the fitting results of circuit model, RC-limited bandwidth can be calculated by Advanced Design System (ADS) software, as shown in Figure 9, where $f_{RC}$ and $f_{3dB}$ denote the simulated RC-limited bandwidth and measured 3-dB bandwidth of the device under 60 μA photocurrent. The value of $f_{RC}$ is much larger than $f_{3dB}$ under low bias from −0.5 V to −3 V, indicating that the overall bandwidth is mainly limited by the carrier transit process in un-depleted absorber. Under −5 V bias, the value of $f_{RC}$ is around twice of the $f_{3dB}$ of the devices with diameter larger than 20 μm, which suggests the RC-limit bandwidth could be comparable to transit time limited bandwidth[28], while the bandwidth of 20 μm diameter device is transit-time limited.

Last but not least, it should be noted that in this work the 3-dB bandwidth of ICIP was characterized by LCA system with 1550 nm wavelength laser. Thus, the carriers excited by incident light are mainly generated in the surface and first stage of the photodetector. This will further cause the current mismatching of each stage in ICIP, which aggravates the high-speed operation and collection efficiency of carriers[16]. Although high reverse bias could partly solve this issue, it is not desirable as it would cause high dark current. One goal of our future work will be using the high-speed MWIR light source to characterize these MWIR photodetectors, which could enable more symmetric absorption profile in each stage, and require less operational bias for high-speed application. Moreover, the device responsivity demonstrated in this work is relatedly low so far. To further improve the responsivity of the ICIP device, multiple stages with current-matched absorber would be desirable, where the number of photo-generated carriers is roughly equal in every stage.

**Conclusion**

In summary, we report a five-stage ICIP based on InAs/GaSb type-II superlattice for MWIR with high frequency operation at room temperature. The 3-dB bandwidth of the five-stage ICIP can achieve up to 7.04 GHz under −5 V bias with low dark current density of 1.04 A/cm[2] at 300 K. According to the analytical model based on S-parameter fitting, this multiple-stage design is mainly limited by diffusion of carriers in the un-depleted absorbers. Future performance improvement would be focused on optimizing the current-matched absorber in multiple stages to boost the overall responsivity of the ICIPs.

AUTHOR INFORMATION

Corresponding Author

Baile Chen − School of Information Science and Technology, ShanghaiTech University, Shanghai, China; Email: chenbl@shanghaitech.edu.cn
Y. Zhou − Key Laboratory of Infrared Imaging Materials and Detector, Shanghai Institute of Technical Physics, Chinese Academy of Sciences, China; Email: zhouyi@mail.sitp.ac.cn

Authors

Zhiyang Xie − School of Information Science and Technology, ShanghaiTech University, Shanghai, China; Shanghai Institute of Microsystem and Information Technology, Chinese Academy of Sciences, Shanghai, China; University of Chinese Academy of Sciences, Beijing, China
Jian Huang − School of Information Science and Technology, ShanghaiTech University, Shanghai, China; Shanghai Institute of Microsystem and Information Technology, Chinese Academy of Sciences, Shanghai, China; University of Chinese Academy of Sciences, Beijing, China
Xuliang Chai − Key Laboratory of Infrared Imaging Materials and Detector, Shanghai Institute of Technical Physics, Chinese Academy of Sciences, Shanghai, China; University of Chinese Academy of Sciences, Beijing, China
Zhuo Deng − School of Information Science and Technology, ShanghaiTech University, Shanghai, China

Yaojiang Chen − School of Information Science and Technology, ShanghaiTech University, Shanghai, China
Qi Lu − Physics Department, Lancaster University, Lancaster, LA1 4YB, United Kingdom.
Zhicheng Xu − Key Laboratory of Infrared Imaging Materials and Detector, Shanghai Institute of Technical Physics, Chinese Academy of Sciences, Shanghai, China
Jianxin Chen − Key Laboratory of Infrared Imaging Materials and Detector, Shanghai Institute of Technical Physics, Chinese Academy of Sciences, Shanghai, China

Author Contributions
[†]Z. Xie, J. Huang and X. Chai contributed equally to this work.
Notes
The authors declare no competing financial interest.

## Acknowledge

This work was supported in part by the National Key Research and Development Program of China under Grant 2019YFB2203400, in part by the ShanghaiTech University startup funding under Grant F-0203-16-002, in part by the National Natural Science Foundation of China under Grant 61975121, 61534006 and 61974152, in part by the Strategic Priority Research Program of Chinese Academy of Sciences under Grant XDA18010000; and in part by the Youth Innovation Promotion Association, CAS under Grant 2016219.

## DATA AVAILABILITY

The data that support the findings of this study are available from the corresponding author upon reasonable request.